# Assessing the quality of home detection from mobile phone data for official statistics

Maarten Vanhoof[1,2], Fernando Reis[3], Thomas Ploetz[1,4], Zbigniew Smoreda[2]

1. Open Lab, Newcastle University, Newastle upon Tyne, United Kingdom.
2. Department SENSE, Orange Labs, Châtillon, France
3. Task Force Big Data, Eurostat, European Commission, Luxembourg, Luxembourg
4. School of Interactive Computing, Georgia Institute of Technology, Atlanta, GA, United States of America

## ABSTRACT

*Mobile phone data are an interesting new data source for official statistics. However, multiple problems and uncertainties need to be solved before these data can inform, support or even become an integral part of statistical production processes. In this paper, we focus on arguably the most important problem hindering the application of mobile phone data in official statistics: detecting home locations. We argue that current efforts to detect home locations suffer from a blind deployment of criteria to define a place of residence and from limited validation possibilities. We support our argument by analysing the performance of five home detection algorithms (HDAs) that have been applied to a large, French, Call Detailed Record (CDR) dataset (~18 million users, 5 months). Our results show that criteria choice in HDAs influences the detection of home locations for up to about 40% of users, that HDAs perform poorly when compared with a validation dataset (the 35°-gap), and that their performance is sensitive to the time period and the duration of observation. Based on our findings and experiences, we offer several recommendations for official statistics. If adopted, our recommendations would help in ensuring a more reliable use of mobile phone data vis-à-vis official statistics.*

## Keywords

*Mobile Phone Data, Home Location, Home Detection Algorithms, Official Statistics, Big Data.*

# 1. Introduction

By now, big data has well and truly arrived and its potential as well as the challenges it poses for official statistics have become much more evident. Consequently, there has been a clear demand to invest in pilot projects that explore how big data can be integrated into official statistics (Eurostat 2014; Glasson et al. 2013).

From this perspective, pilot projects are useful not only to identify practical issues (e.g. legal issues, data-management) but also, they are particularly useful when critically assessing the reliability of data sources and methodologies. It is reassuring that, regarding such assessments, Karlberg et al. (2015, p.1) observe: "There is a clear trend towards a more reflective approach, with an emphasis not only on producing high-quality statistics, but also on rendering explicit details on exactly how this is being achieved". However, the importance of providing explicit details is not to be underestimated when it comes to big data: big data sources, typically, do not adhere to official statistics' standards and principles - such as issues on coverage, representativity, quality, accuracy and precision (Daas et al. 2015) - and, consequently, neither do their methodologies.

In this paper, we conduct a pilot study focusing on, arguably, the most important step for the application of mobile phone data in official statistics: identifying where someone lives, this is, detecting the home location. Current home detection methods for mobile phone data do not adhere to official statistics' standards (or even to what could be reasonably expected from academic standards). We elaborate our argument by means of an extensive review of literature and an empirical analysis based on a large-scale, French, Call Detailed Records (CDR) dataset. In doing so, we aim to show how current home detection practices came to be, how they are bound by limited validation possibilities and how they are sensitive to criteria choice or decision rule development. Given the lack of research on these problems, we argue that there is no clear framework on which to apprise the performance or the uncertainty of current home detection methods.

Our analysis evaluated the performance of five different home detection algorithms using a mobile phone dataset from France. The case study allows us to reflect on the findings from a more practical point of view, whilst also contributing to our discussions and recommendations on the various uncertainties that underlie current home detection practices. We hope our contribution will help other researchers and practitioners to recognise the difficulties of integrating information on home locations sourced from mobile phone data into official statistics.

## 2. Mobile phone data, official statistics and the role of home detection methods

### 2.1 Mobile phone data and official statistics

Before looking at the methods used to identify home location, let us quickly consider how mobile phone data can be of interest for official statistics.

Over the last decade, the analysis of this type of data source has grown into a mature research field and, based on findings, applications are being developed and applied (Blondel et al. 2015). One line of interest is that mobile phone data has the potential to capture temporal patterns of user presence (Deville et al. 2014), which could be used to estimate population density (Ricciato et al. 2015). In turn, these estimations could usefully support official statistics in developing countries (Blondel et al. 2012; de Montjoye et al. 2014).

Another line of interest relates to the large-scale recording of mobility patterns. As mobile phones can capture individual mobility for millions of users, applications have been developed that estimate nationwide commuting figures (Kung et al. 2014), long-distance trips (Janzen et al. 2016, Janzen et al. 2018), inbound tourism trips (Raun et al. 2016) and even domestic tourism trips (Vanhoof et al., 2017b).

These and similar developments have the potential to enhance official statistics in fields such as the delineation of urban areas (Vanhoof et al. 2017a), the understanding of migration patterns (Blumenstock 2012), or to complement tourism statistics (Ahas et al. 2008). They could even perform *nowcasting* of macro-economic and socio-economic aspects of populations (Baldacci et al. 2016; Marchetti et al. 2015; Giannotti et al. 2012; Pappalardo et al. 2016, Vanhoof et al., 2018).

### 2.2 The role and method of home location

Common to many, if not all, mobile phone data research is the need to identify the home location of mobile phone users before proceeding to more advanced analysis. For example, knowing the place of residence is a prerequisite before analysing the amount of time spent at home and commuting patterns, which in turn fuel mobility and epidemiological models (Rubrichi et al. 2018). Besides its relevance within mobile phone data analysis, knowledge of home location also forms the crucial link between mobile phone data and other data sources such as census data, making it a key enabler for the combination of information.

The method of pinpointing where someone lives consists of attributing a supposed home location to every single user in the database from the geographical metadata obtained from their mobile phone records. In practice, identifying a person's home means that a single cell tower is allocated as their home location. This allocation is based on the calling and movement patterns of each individual user. The spatial resolution of cell towers is used because most mobile phone datasets only have geographical data for the towers' positions. The assumption then is not that a user lives at that exact cell tower location, but rather somewhere in the area covered by the tower. It is remarkable that even though detecting the home location now forms a cornerstone of mobile phone research, home detection methods are often obscured in literature: details on their exact application, related uncertainties, perceived performance or even the validation processes are only rarely communicated.

In the following section, we show why current home detection practices are problematic. In an extended literature review, we show how, over time, methodologies for home detection have been simplified to single-step approaches using decision rules that are based on simple, *a-priori* defined criteria of what defines a 'home'. Such methods are questionable because of the possibilities to validate are limited, plus there is a lack of knowledge on their sensitivity, specifically in respect to criteria choice. Our empirical work with a large, French, mobile phone dataset exemplifies several of the problems we raise. It allows us to put the problems in a more practical context and to outline their consequences in more detail.

## 3. Identifying homes from large-scale location traces

Given the enormity of the datasets that capture geo-located traces of users, literature explains the automated methods developed for identifying the homes, or other meaningful places such as the workplace, of users. Here, it is necessary to distinguish between continuous location traces (e.g. GPS data) and non-continuous location traces (e.g. mobile phone data) where the latter do not provide a similar high-volume, high-resolution capture of location traces in time or space compared to the former.

As our main interest is to outline the deficits in the methods used for non-continuous location traces, this section will start by reviewing the literature on automated home location.

## 3.1 Identifying meaningful places from continuous location traces

The analysis of continuous location traces has been the focus of early developments in the automated identification of meaningful places. Related work typically used small-scale datasets, most commonly from continuous GPS traces but also from Bluetooth, or Wi-Fi positioning (Wolf et al. 2001; Shen and Stopher 2014). The general methodology used to identify meaningful places from continuous location traces, consists of a two-step approach.

In the first step, location traces are clustered in space (and sometimes in time) in order to detect *important places*. Techniques for clustering continuous location traces range from manual GIS analysis (Wolf, Guensler, and Bachman 2001; Gong et al. 2012) to automated, unsupervised analysis using, for example, k-means clustering (Ashbrook and Starner 2003), non-parametric Bayesian approaches (Nurmi and Bhattacharya 2008), or fingerprinting of the radio environment (Hightower et al. 2005).

In a second step, the important places identified are then annotated as *meaningful places* (such as home, work, recreation area). Annotation can be done either through interpretation, for example by expert judgment, by surveying the user that produced the traces, or through automation, mainly by means of time-space heuristics (Nurmi and Bhattacharya 2008).

## 3.2 Identifying meaningful places from non-continuous traces

In contrast to the above-mentioned continuous location traces, the use of non-continuous location traces has recently become very popular. Examples of activities that produce non-continuous location traces in large-scale datasets are mobile phone usage, credit card transactions, or check-ins through location based services (e.g. Foursquare) and online social networks (e.g. Twitter).

The identification of meaningful places from non-continuous location traces poses substantial challenges, most notably due to the less frequent observations and the larger spatial resolution in which observations are captured (e.g. mobile phone data are only captured at the location of the cell tower used). These challenges, however, are outweighed by the presumed advantages associated with the larger coverage, in terms of users, timespan and spatial extent of the data sources (Järv, Ahas, and Witlox 2014; Kung et al. 2014).

The following analysis will focus on one example of how to identify one meaningful place - the location of a user's home, using one prominent example of non-continuous traces: Call Detailed Record (CDR) data.

CDR data are mobile phone data captured by the network operator every time a user makes or receives a text or call (hence the non-continuous tracing). Note that the methods and problems described in next sections are not limited to CDR data, but are relevant for all datasets covering non-continuous location traces.

### 3.2.1 Two-step approaches for non-continuous traces

As with the two-step approaches for continuous traces, initial methods to detect home locations from CDR data also clustered location traces into important places before annotating them as meaningful places. For example, in Isaacman et al. (2011) individual traces from CDR data are clustered using Hartigan's leader algorithm. Clusters are then annotated into meaningful places by means of a logistic regression model that is trained on data from 18 persons for which ground truth was available. Next, and for each user, the cluster with the highest score on the logistic regression model is chosen to be the presumed home area.

### 3.2.2 Single-step approaches for non-continuous traces

However, two-step approaches for non-continuous location traces quickly gave way to single-step approaches that are now widely deployed in literature (Calabrese et al. 2014; Calabrese et al. 2011; Kung et al. 2014; Phithakkitnukoon et al. 2012). The difference between two-step and single-step methods is that the latter skips the clustering into important locations and thus acts directly on individual cell towers instead of groups of cell towers.

One of the reasons for switching to single-step approaches is that the standard clustering methods used in the two-step approaches make it difficult to construct consistent spatial traces when combined with non-continuous location traces. Nevertheless, the main drawback of this switch to a single-step approach is that the spatial pattern of the location traces is largely neglected, as only single cell tower annotation is targeted. This increases the uncertainty of fixing home location because single events at individual cell towers may be sufficient to undermine the method.

In practical terms, detecting a home in a single-step approach is done by using a decision rule that is based on an *a-priori* definition of home - the *home criterion* as we call it - in order to produce a list of one or several cell towers that could be the home location. A standard example of a home criterion for the case of CDR data is: 'home is where calls are made during the night'. The problem with single-step approaches is that such decision rules are being applied as *heuristics*, meaning that one general rule is applied to the location traces of all users even though a different set of decision rules could potentially lead to better results.

In terms of identifying home location, applying heuristics implies that meaningful places (like the home) can be described similarly for all users in the dataset, regardless of the user's characteristics as observed in their movements and calling patterns. It seems logical that the imposition of this assumption can only be done when a proper evaluation and validation of their movements has been carried out, or when clear evidence exists for the use of a specific criterion or decision rule. For this reason, the following paragraphs will discuss how to define decision rules for one-step home detection methods and which criteria to use.

### 3.3 Defining decision rules for single-step home identification

*3.3.1 Simple decision rules for single-step home detection*

The core challenge for single-step home detection is in defining a decision rule that is simultaneously capable of i) distinguishing between different important places and ii) annotating the correct home location. Most research employs simple decision rules that are either based on information from official statistics or rely on precedents found in literature.

When examining the existing decision rules in research literature, the most popular are: time-based limitations for the night ('home is the location that has the most activity between x pm and y am'), time-based aggregations ('home is where the most distinct days, or weekend-days are spent) as well as spatial groupings ('home is the location with the most activity in a spatial radius of x km around it), (Calabrese et al. 2011; Phithakkitnukoon, Smoreda, and Olivier 2012; Frias-Martinez and Virseda 2012; Kung et al. 2014; Tizzoni et al. 2014). One example, using time-interval statistics from a Boston dataset drawn from the American Time Use Survey (Calabrese et al.2011), uses the highest distinct number of observations between 6pm and 8am to derive home locations.

Almost all studies using simple decision rules rely on census data. They depend either on specific surveys and questionnaires to define the criteria deployed, (Calabrese et al. 2011) or, for high-level validation, on aggregated population density data (Phithakkitnukoon et al. 2012) or commuting figures (Kung et al. 2014).

### 3.3.2 Complex decision rules for single-step home detection

A few studies have elaborated more complex decision rules for home detection. The seminal work of Ahas et al. (2010), for example, uses a tree-based approach that combines a set of criteria including distinct days of activities on a cell tower, the starting times of calls, deviations of starting time of calls, durations of calls, and this all for a training set of 14 people for which the ground truth was known. The decision rules, as defined by the classification tree, were consequently deployed to all users in an Estonian dataset (as heuristics in other words), raising the question of how representative a training set of 14 people could possibly be for a large population.

The problem of small training sets was overcome in Frias-Martinez et al. (2010) who used a training set of 5000 users to construct a complex decision rule for home detection. Deploying a Genetic Algorithm technique, they focus on finding the best combination of temporal criteria to denote home locations in an emerging economy. Their best performance is a correct prediction of around 70% for a subset of 50% of the users. Users were filtered on the basis of having at least a 20% difference in the percentage of total calls between the first and second eligible cell tower. The complex decision rule they use to obtain this result is to select the cell tower logging the most activity during the nights of Friday, Saturday, Sunday, Monday and Tuesday from 5:15 pm to 8:30 am. The individual ground truth data in Frias-Martinez et al. (2010) are retrieved from users' contracts with the provider. This data is not available in most countries due to legal obligations to anonymise users or to bans on linking individual information to CDR data.

As a consequence, Csáji et al. 2013, try to derive a temporal decision rule but this time without a training/validation dataset at individual level. Applying an unsupervised k-means algorithm to the temporal activity patterns of frequently used cell towers in Portugal, they find clusters that are interpretable as temporal patterns typically relating to presence at home, at work, or as not interpretable at all. Consequently, their decision rule to detect home locations is based on these temporal patterns interpreted as home presence. Compared to Frias-Martinez et al. (2010), one of the drawbacks of their approach is that they did not construct their criteria based on individual observations. This raises the question as to the degree to which such criteria are realistic for different subsets of users.

In a way, the subset representativity problem persists for all single-step approaches, regardless of whether their decision rules are defined in a complex or simple way. If the same decision rules is applied to all phone users, careful investigation into the effect at individual level, or at population subset level should be carried out, in order to know the degree to which generalisation favours or disfavours subsets of users. In other words, if decision rules are applied generically, in-depth validation of the single-step approaches is important.

### 3.4 Validating large-scale home detection methods

The use of a particular decision rule, whether derived from a census, borrowed from literature or defined by training sets, is often based on comparing population counts from mobile phone data with census data. However, such high-level validation does not offer a direct evaluation of performance at individual user level, nor does it allow for comparison between cases. In fact, assessing the performance of different decision rules by comparing the resultant population counts with census data is, strictly speaking, a rather limited alternative solely justified by the absence of individual level validation data.

The absence of validation data at individual level is a common problem in published research, and is therefore often taken for granted. But the absence of validation data has several consequences. First and foremost, it impedes the creation of evaluation metrics that can assess the performance of home detection at individual level. Such an individual level evaluation could allow us to better understand the workings of different decision rules on a specific dataset and user subsets, which in turn could enable a comparison between different decision rules, datasets, users and areas.

Secondly, the absence of validation data at individual level is implicitly why single-step approaches apply decision rules as heuristics. In the absence of individual level validation data, it is impossible to understand which decision rules works best for any individual user. Consequently, case-adjusted, adaptive algorithms cannot be developed. This implicitly forces researchers and practitioners to adhere to a one-size-fits-all solution in order to be clear and consistent.

It is worth noting that, currently, high-level validation is still assumed to be a good solution in the absence of individual level validation data. In particular, two observations stand out.

Firstly, census data is often used for high-level validation. For example, comparisons for small geographical areas can be made between the counts of home locations identified from mobile phone data and the aggregated counts of peoples' residential locations obtained from censuses. This is a very opportunistic, if not naïve, validation attempt as census data has never specifically been gathered to serve this purpose and little or no information exists on how, for example, different spatial delineations or the distorted market shares of mobile phone operators could influence this kind of validation.

Secondly, it is noteworthy that no studies have used high-level validation to compare the performance of different decision rules. Nor are there studies that evaluate the sensitivity of high-level validation to criteria choice. This absence is probably because high-level validation is not informative enough to properly understand the differences between criteria, decision rules, and their performances. Given this, we are far from obtaining a consensus on which criteria are best, or on how to construct optimal decision rules. In fact, we are far from understanding the strengths and weaknesses of different home detection methods altogether. Given this, we should question the degree to which high-level validation contributes to the development and trustworthiness of home detection.

### 3.5 Current deficits of home detection using non-continuous location traces

In conclusion, we find a clear framework is missing to allow us to understand the performance, uncertainty and sensitivity of the criteria choice or decision rule development, especially at individual level, when using non-continuous location traces to detect home location. Despite their widespread use, no clear reasoning exists as to why single-step approaches should be chosen over two-step approaches. Nor does a consensus exist on which criteria should be used, or how optimal decision rules for a given dataset should be defined.

Similarly, it is striking that no work investigates the sensitivities of single-step approaches to criteria choice. Additionally, we find that the validation of large-scale home detection methods is severely limited because of the absence of ground truth data at individual level. As a result, current assessments of home detection methods are based on high-level validation, but the trustworthiness and exact contribution of this practice is rather dubious.

In summary, our findings indicate that the current methods to identify users' home locations for official statistics are rather questionable. We illustrate some of the aforementioned problems by means of a case study for identifying home locations using French CDR data.

## 4. Investigating home detection algorithms for French CDR data

To explore the application of single-step home detection methods on a French CDR dataset, we start by constructing five home detection algorithms that incorporate different popular home criteria in simple decision rules. We apply these algorithms to the French dataset, perform high-level validation, and investigate sensitivity to criteria choice. This allows us to demonstrate some of the aforementioned problems in an applied context.

## 4.1 The French CDR dataset

CDR data are the most widely-used examples of mobile phone data in research. CDR data are passively gathered by operators for billing and maintenance purposes and are collected every time a mobile phone user makes or receives a text or a call. Apart from technical metadata on the workings of the network, CDR data contain information on the time, the location (the cell tower used), as well as the caller and the call receiver.

For our analysis, we use an anonymised CDR dataset from the mobile phone carrier Orange. The data covers the mobile phone usage of ~18 million users on the Orange network in France during a period of 154 consecutive days in 2007 (May 13, 2007 to October 14, 2007). At that time, mobile phone penetration was estimated at 86% (ARCEP 2008). Given a population of 63.9 million inhabitants during the observed period[2], this dataset covers about 32.8% of all French mobile phone users and 28.6% of the total population.

The Orange France 2007 CDR dataset is one of the largest CDR datasets available in terms of population-wide coverage and has been extensively studied before (Grauwin et al. 2017; Sobolevsky et al. 2013; Deville et al. 2014). It is the latest CDR dataset available for France that allows for long-term, temporal continuous tracking of mobile phone users. Access to more recent datasets is limited by The French Data Protection Agency (CNIL), which is anticipating the EU General Data Protection Regulation and does not allow individual traces for periods of more than 24 hours to be collected, before being irreversibly recoded.

Some of the typical characteristics of CDR datasets that pose substantial challenges for their automated analysis, are the temporal sparsity in observations (only a few records per user per day[3]), and the spatially uneven distribution of the areas covered. The latter results from a demand-driven, non-uniform distribution of cell tower locations[4] (higher densities of cell towers are found in more densely populated areas such as cities or coastlines).

On the other hand, it is very attractive to have the possibility of researching the large-scale CDR datasets at population level, without users needing to share their locations. This increases the feasibility of automated applications such as home location. In addition, continuous data collection allows us to observe over extended periods, which in turn enables complex analysis and lessens any influence emanating from singular events and/or non-routine behaviour.

---

[2] The average of monthly estimates between May and October 2007 as obtained from the INSEE Website (www.insee.fr).

[3] For example: for an arbitrary day of the timespan covered (Thursday, 1st October 2007), the median number of records per user was four, relating to only two different locations. Such statistics are representative for CDR based studies and can be deemed rather high compared to other large-scale non-continuous datasets like credit-card transactions or Flickr photos (Bojic et al. 2015).

[4] The spatial accuracy of the dataset is restricted to the network's spatial resolution. In France, the distribution of 18 273 cell tower locations is known but is not uniform.

## 4.2 Applying five HDAs to the French CDR data

*4.2.1 Constructing five HDAs with simple decision rules based on popular home criteria*

To perform home detection, we construct five basic Home Detection Algorithms (HDAs). Each incorporates one or two popular home criteria that are applied by means of simple decision rules. In order to select criteria, we took into account literature that dealt with single-step approaches (a.o. Ahas et al. 2010; Isaacman et al. 2011; Calabrese et al. 2011; Tizzoni et al. 2014; Chen, Bian, and Ma 2014; Phithakkitnukoon, Smoreda, and Olivier 2012; Csáji et al. 2013; Kung et al. 2014). We also used distilled criteria that were sometimes used independently (e.g. Tizzoni et al. 2014), sometimes combined (e.g. Ahas et al. 2010), sometimes within simple decision rules (e.g. Phithakkitnukoon, Smoreda, and Olivier 2012), and sometimes within complex decision rules (e.g. Csáji et al. 2013 and Frias-Martinez 2010).

The HDAs we construct use the decision rules that 'home' is in the area of the cell tower where:

1. The majority of both outgoing and incoming calls and texts were made (amount of activities criterion);

2. The maximim number of distinct days with phone activities —both outgoing and incoming calls and texts— was observed (amount of distinct days criterion);

3. Most phone activities were recorded during 7pm and 9am (time constraints criterion);

4. Most phone activities were recorded, implementing a spatial perimeter of 1000 meters around a cell tower that aggregates all activities within (space constraints criterion); and

5. The combination of 3) and 4), thus most phone activities recorded during 7pm and 9am and implementing a spatial perimeter of 1000 meter (time constraints + space constraint criterion).

Note that throughout this paper, we will estimate cell tower areas by means of the Voronoi tessellation of the cell tower network. The use of Voronoi polygons to describe the spatial patterns of cell tower coverage has disadvantages. Although widely used in literature, Voronoi polygons are a simplification of the actual capacity of cell towers to cover areas. In reality, capacity is dependent on factors such as humidity, urban environment, elevation of the cell tower, and orientation. Theoretically, developing estimation models for the coverage of cell towers should be possible, but such models need extensive field surveys for validation, surveys that are expensive and thus rarely available. Therefore, there exists an unobservable measurement error when using Voronoi polygons and most findings (including ours) are dependent on the assumption that this error has an insignificant impact.

Note also that Bojic et al. (2015) uses similar HDAs when assessing and comparing home detection methods for a credit card transaction and Flickr dataset. This shows that the relevance of these algorithms goes beyond the case of CDR data and also serves other datasets with non-continuous location traces.

*4.2.2  Applying five HDAs to the French CDR dataset.*

We apply all five HDAs to the Orange France 2007 CDR dataset, to detect the cell tower that covers the presumed home location (L1) for all users during all months in the dataset (May-October). Besides the L1 cell tower, we gather information about the second (L2) and the third (L3) most plausible cell tower to cover the home location following the particular decision rule applied.

Table 1 shows the total number of times each HDA could detect an L1, L2 or L3 cell tower based on the CDR data of ~18 million users and when applied to each month in the dataset. Given the availability of six different months (mid-May to mid-October), non-restrictive algorithms (such as algorithm 1 and 2) will be capable of detecting an L1 cell tower for about 109.4 million users (~18 000 000*6). Restrictive algorithms, such as the time constraining algorithm 3, have fewer users for which a presumed home cell tower (L1) can be detected. The reason is that some users might not have made or received calls or texts during the restricted timeframe, so no CDR records exist and therefore the algorithm cannot identify an L1 cell tower.

Table 1: Number of times (in millions) for which an L1, L2 or L3 cell tower could be detected from an individual user's CDR data by the various Home Detection Algorithms (HDAs) when applied per month in the dataset. Percentages are column-wise and with respect to the number of L1 detections.

| Number of users (in million) with | Amount of activities (algorithm 1) | Amount of distinct days (algorithm 2) | Time restraints (amount of act). (algorithm 3) | Space restraints (amount of act.) (algorithm 4) | Time and Space restraints (algorithm 5) |
|---|---|---|---|---|---|
| Detected L1 | 109.4 (100%) | 109.4 (100%) | 98.4 (100%) | 109.4 (100%) | 98.4 (100%) |
| Detected L2 | 102.2 (93.5%) | 102.2 (93.5%) | 78.0 (81.3%) | 102.0 (92.8%) | 78.4 (79.6%) |
| Detected L3 | 96.1 (87.9%) | 96.1 (87.9%) | 65.0 (66.1%) | 66.1 (86.6%) | 62.3 (63.3%) |

For example, when we compare the number of times algorithm 1 (all activities) was capable of detecting an L1 compared to algorithm 3 (only nighttime activities), we can derive that up to 10% (98.4/109.4) of the users did not have mobile phone activities during the night. This made it impossible for the time-constraint HDAs to detect a cell tower presumably covering the home location. It is also interesting to note that, depending on decision rule of the algorithm, between 79.6 and 93.5% and between 62.3 and 87.9% of users have an L2 or L3 cell tower that could also be nominated as the home location cell tower, as they only varied by a slight degree compared to the L1 (or L2) cell tower(s). In other words, the decision rules applied do not overly discriminate between the eligibility of different cell towers to be the presumed home location. This raises the question of whether the French dataset would not have benefited from a two-step approach.

### 4.3 Comparison of HDAs at individual level

One intriguing question is whether, for the same individual user, different HDAs would detect different home locations (L1 cell towers). We assess to which degree two different algorithms detect similar home locations for all individual users in the dataset by evaluating the Simple Matching Coefficient (SMC) (Bojic et al. 2015):

$$\%SMC\ (algorithm_A, algorithm_B) = 100 * \frac{\sum_{i=1}^{N} \delta(Home_{A,i}, Home_{B,i})}{N} \quad (1)$$

where $i=1..N$ denotes the N users analysed, and $\delta(Home_{A,i}, Home_{B,i})$ is the Kronecker delta which is equal to 1 when the home detected by algorithm A for the $i$-th user is identical to the home detected by algorithm B for the same user. The Kronecker delta becomes 0 otherwise. Values of %SMC thus range between 0 and 100 and can be interpreted as the percentage of individual cases for which both algorithms detected the same home locations. When calculating SMC values, we omit all cases where one of the algorithms failed to detect a home location (e.g., when no observations were left after implementing a time constraint).

Figure 1 shows the SMC values for all pair combinations of HDAs during the different months in the dataset. In general, pair accordance ranges between 61.5% and 96.4% of the detected homes, resulting in discordance rates between about 40% and 4%. In absolute numbers this means that different decision rules predict different homes for between 6.8 and 0.6 million users. The patterns of (dis)similarities between HDAs are rather clear. Algorithms that incorporate time-constraints (algorithms 3 and 5) have a high degree of variance with algorithms that count the amount of activities (algorithm 1), distinct days (algorithm 2), or perform spatial groupings (algorithm 4), all of which show rather high degrees of pair accordance. The different results for the time-constraints algorithms might stem from sparser observations or different movement patterns during the night, but exact reasons are unknown.

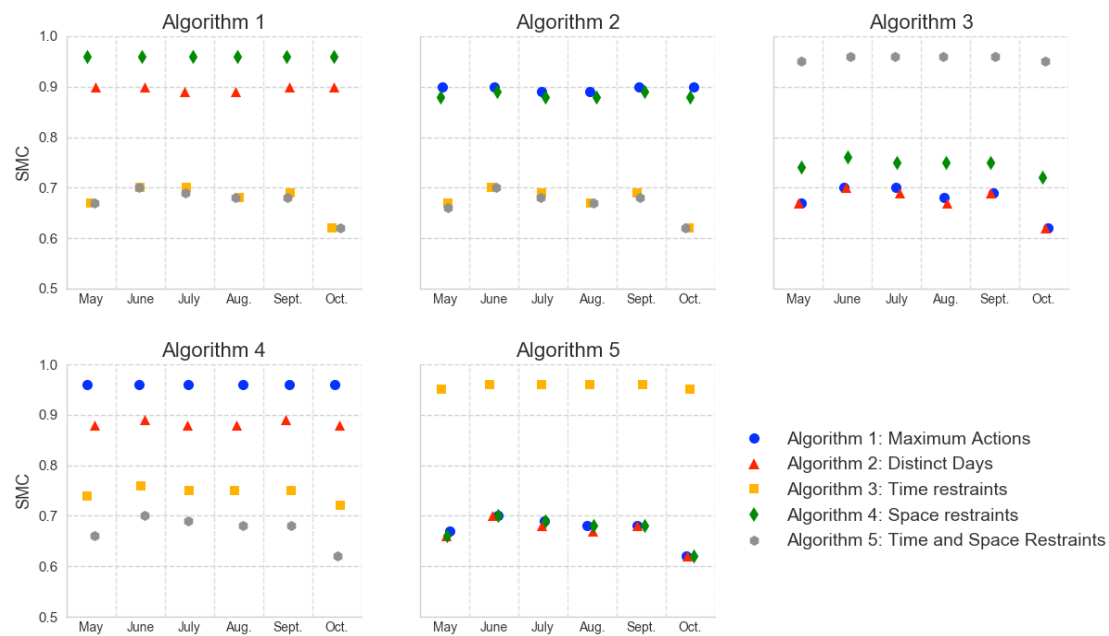

Figure 1: SMC values for all pair combinations of HDAs, for each month in the dataset. SMC values express the ratio of users for which two HDAs detect the same home.

## 4.4 High level validation of Home Detection Algorithms

Given that different HDAs give different results for a considerable share of all individual users, the question becomes which decision rule should be preferred. As discussed previously, no consensus exists in literature on which decision rule(s) are best. This is partly because of the absence of comparative studies, but mainly because of the lack of proper validation data at individual level. In our case too, individual-level ground truth data was not available and so our assessment is at high-level, comparing census figures with population counts produced by HDAs.

*4.4.1 National Statistics Validation Dataset*

In contrast to related works, our high-level validation is based on a unique validation dataset that was created in collaboration with the French National Statistics Institute (INSEE). To construct the validation dataset, the Public Finances Directorate General (DGFIP) collected individual (or household) home locations from revenue declarations, housing taxes and the directory of taxable individuals. It then aggregated this information into population counts at the resolution of the Orange cell tower network (see also Figure 3a). In other words, an estimation of the population numbers, based on census data, for the geographical areas created by the Voronoi polygons of the Orange cell towers was produced and made available to the research project under a non-disclosure agreement.

It is a huge advantage to have access to a validation dataset which has the same spatial resolution as the mobile phone network. It avoids the spatial translation of statistical zones to the cell tower Voronoi areas, which is complicated and prone to errors (Frias-Martinez et al. 2010), given the spatially uneven distribution of cell towers.

Unfortunately, the individual (or household) home locations used to construct the validation dataset could only be made available for the year 2010. However, for reasons explained in the previous paragraph we do opt to use this validation dataset with its temporal mismatch (the mobile phone dataset covers 2007) over the low resolution, publicly available census data that are updated every year. Since we only use the validation dataset for relative comparisons between HDAs (i.e. no absolute validation is attempted), the assumption we introduce concerning this temporal mismatch is that relative population patterns do not change drastically within three years.

*4.4.2   Validation of HDA Results at Cell Tower Level*

To compare results from HDAs with the proposed validation dataset, we evaluate the degree of similarity in population counts attributed to all cell tower areas. Note that we do not target an absolute assessment of similarity, as this is impossible given the unknown spatial distribution of the 28.7% sample of Orange users and the differences in times of collection between the CDR dataset (2007) and our validation data (2010). Instead, we compare general patterns of estimated populations by means of vector comparison.

In our case, a first vector denotes the estimated population by one HDA for all cell tower areas and is compared to a second vector that describes the validation population count for exactly the same cell tower areas. Both vectors thus have an equal length (*n*=18 273 the amount of cell towers in the Orange network). To quantify the similarities and differences between both vectors, we use a standard Cosine Similarity Metric[5] (CSM), which is based on the angle between two vectors described by its cosine:

$$\cos(\vec{x}, \vec{y}) = \frac{\sum_{i=1}^{n} x_i * y_i}{\sqrt{\sum_{i=1}^{n} x_i^2} * \sqrt{\sum_{i=1}^{n} y_i^2}} \quad (2)$$

where $x_i$ and $y_i$ are components of vector $\vec{x}$ and $\vec{y}$ respectively and *n* is the total number of cell tower areas.

Values of the cosine will range between -1 and 1. A value of 1 indicates the highest similarity in orientation (the angle between $\vec{x}$ and $\vec{y}$ is zero degrees), 0 indicates the lowest similarity in orientation (the angle between vector $\vec{x}$ and vector $\vec{y}$ is 90 or -90 degrees) and -1 indicates an opposite orientation (the angle between $\vec{x}$ and $\vec{y}$ is 180 degrees). Deriving the angle between two vectors and expressing it in degrees (°) consequently gives us the CSM value we want:

$$\text{CSM}(\vec{x}, \vec{y}) = \left| \cos^{-1}\left(\frac{\sum_{i=1}^{n} x_i * y_i}{\sqrt{\sum_{i=1}^{n} x_i^2} * \sqrt{\sum_{i=1}^{n} y_i^2}}\right) * \frac{180}{\pi} \right| \quad (3)$$

A CSM value of 0° denotes the highest possible similarity, 90° indicates the lowest similarity in orientation whereas 180° degrees refers to an opposite orientation.

---

[5] According to Ye (2011 p.91): 'The cosine similarity is a classic measure used in information retrieval and is the most widely reported measure of vector similarity'. It is, for example, defined in the seminal work of Salton and McGill (1986): Introduction to Modern Information Retrieval.

*4.4.3 Validation with census data: CSM*

Figure 2 shows the calculated CSM values for all HDAs and for different months. The distinct days algorithm performs best in replicating the population pattern of the validation dataset, followed by the number of activities and the time-constrained number of activities. The HDAs that involve grouping in space perform worst, even though the applied perimeter (1 kilometre) in reality does not correspond to a substantial distance. It is worth noting that the performance of all HDAs range between 34° and 38°. This is substantially different from the intended 0°, which would signify a perfect match with the validation set. In other words, a 'gap' of about 35 degrees exists when using the CSM measure. This is indicative for the limited performance of our HDAs and raises the question of whether there is a structural limitation on the performance of single-step HDAs when applied to the French dataset or to CDR data in general.

Interestingly, the performance of all HDAs is rather similar. Especially in their temporal patterns where lower CSM values for June and September, and higher values for May, July, August and October are observed. A possible explanation for the high SMC values for May and October is the limited number of available days for these months in the dataset (18 and 14 days respectively). This indicates that data should be collected for a certain duration for the HDA to perform properly.

The highest CSM values are observed during summer (July and August). All algorithms are sensitive to this temporal change, most likely because of the changing spatial behaviour of users who go on holiday (see also Deville et al. 2014; Vanhoof et al., 2017b). Time-limited criteria are more sensitive to temporal changes, which raises questions about their widespread adaptation in literature. In addition, it is interesting to note that differences between each algorithm are smaller than the differences of each algorithm over time. Future analysis of HDA performance should therefore take into account the time period.

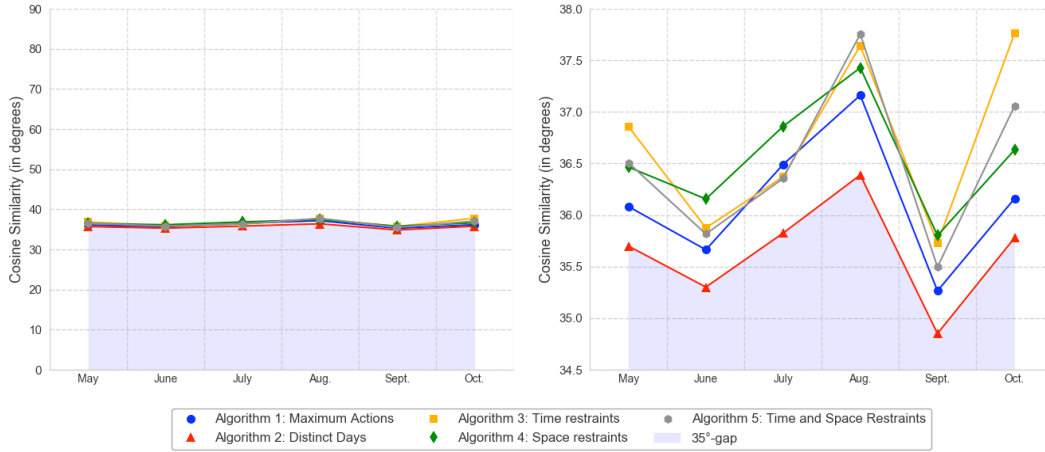

Figure 2: CSM values (in degrees) of the comparison with ground truth data, for all HDAs applied to all months in the dataset. CSM values were calculated at cell tower level. The 35°-gap is denoted as the difference between the best performing HDA and the expected CSM of 0° in the case of a perfect match between population counts from home location and the validation dataset.

*4.4.4. Spatial patterns of population count*

Although the CSM values for all HDAs are within a rather small range, it is important to realise that small differences in CSM values can imply quite major differences in the related spatial patterns of population counts.

Figures 3c and 3d for instance, show the spatial patterns of population counts obtained by the number of activity algorithms for June and August respectively. The difference in CSM values between both is a mere 1.08° but their spatial pattern, as emphasised by the Getis-Ord $G_i^*$ statistic (Getis and Ord 1992), is rather different. This statistic shows statistically significant clusters of high (hotspots) or low (coldspots) population counts. In August, for instance, the detected hotspots illustrate clear clusters of high numbers of home locations near sea and mountain areas. This is in contrast to an expected spatial pattern, where high clusters of population counts are found near cities and in urban areas, as can be seen from the spatial pattern of the validation dataset in Figure 3b.

The spatial pattern of the differences between the validation datasets and detected homes in June and August are given in Figures 3e and 3f and visualise this contrast. Note that in Figure 3, the centre of Paris is often denoted as a coldspot because of the high density of cell towers, so each tower has a lower number of users, resulting in apparent coldspots. This effect is also visible for other city centres where cell tower density is high.

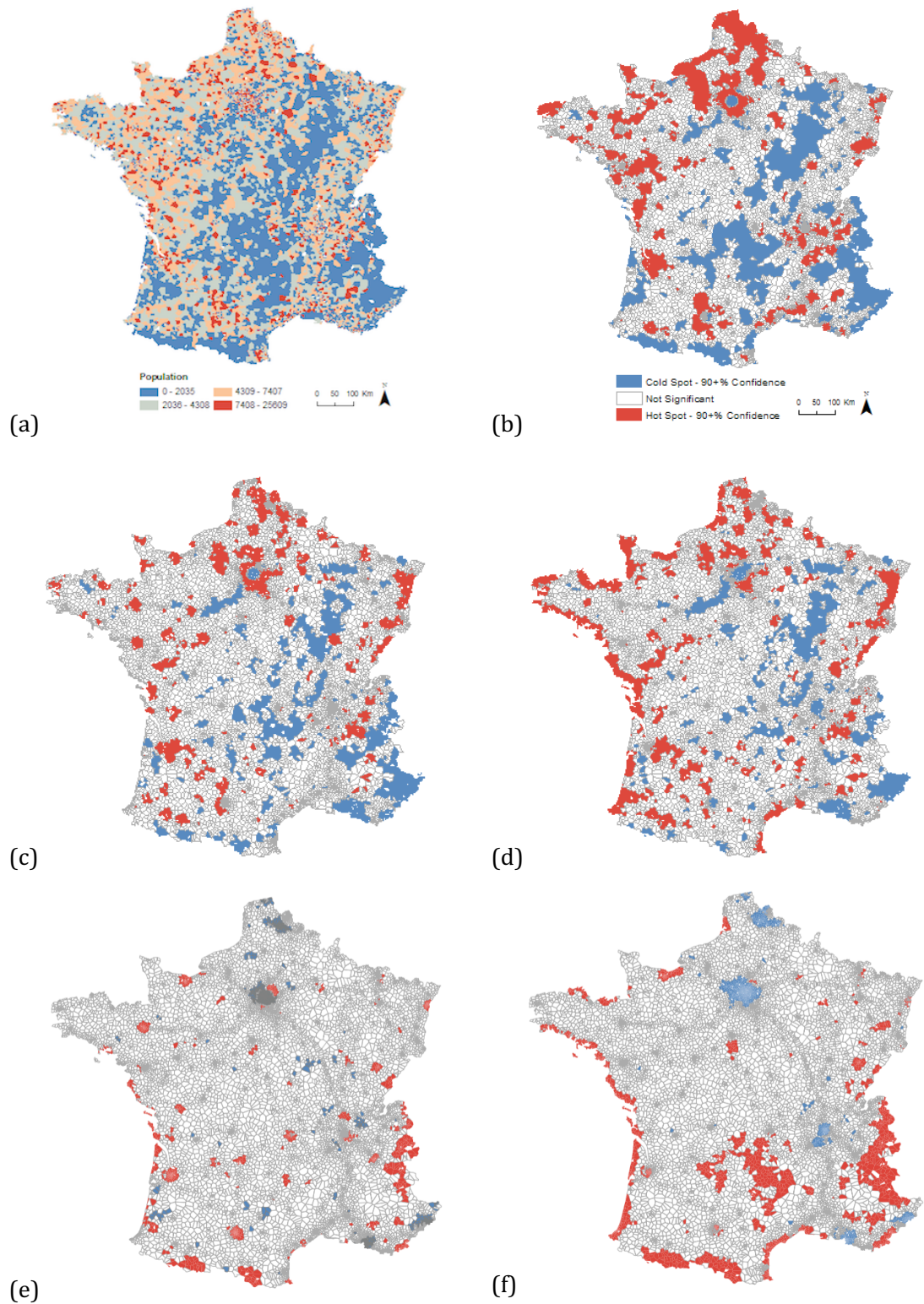

Figure 3: Population counts of the validation dataset: (a) Hotspots (red) and coldspots (blue) as defined by the 90+% interval of the Getis-Ord $G_i^*$ statistic for the population counts of the validation dataset (b); for the number of detected homes using the amount of activities algorithm in June (c); for the number of detected homes using the amount of activities algorithm in August (d); for the log(ratio) between the amount of activities algorithm in June and the population counts of the validation dataset (e); and for the log(ratio) between the amount of activities algorithm in August and the population counts of the validation dataset (f). All maps are compiled from Voronoi tessellation of the Orange cell towers. Figures b, c, d, e, and f share the legend of Figure b.

# 5. Discussion

## 5.1 Differences at individual level and the absence of ground truth data

Our results showed high discordance rates between different HDAs (ranging from 4% to 40% of the individual mobile phone users). This finding challenges the use of single-step home detection approaches for the French CDR dataset when done without fully justifying the home location criteria used and the decision rules involved in the HDAs. As we argued, such justification is currently absent in the aforementioned literature mainly because of the absence of ground truth data at individual level. Our case study clarifies how the absence of individual ground truth data necessitates a heuristic application of decision rules in current home detection methods. By this we mean that one decision rule is applied to all users in the dataset, regardless the nature of their CDR traces. Obviously, the better approach would be to have non-generic algorithms that could flexibly select decision rules (and validation) based on the characteristics of individual user traces. Such a solution, however, would require large training samples (individual ground truth) to learn how to switch between different decision rules. As yet, these are not available.

## 5.2 Sensitivity of performance considering time and decision rule choice.

Performing high-level validation on five HDAs, by comparing population counts with census data, unveiled rather poor performances (CSM values between 34° and 38°), and a clear sensitivity to the chosen time period. In fact, for the French dataset, defining a time period to carry out home detection seems as important as criteria choice. As was illustrated by the spatial pattern of population counts in Figure 3d, and by increasing CSM values in July and August in Figure 2, summer periods should best be avoided when running HDAs. Additionally, shorter observation periods (like May and October in our case) also seem to influence the performance of HDAs.

When comparing criteria, it is clear that the space constraints criterion, is outperformed by all others. The main logic behind grouping close locations together (in this case, within a 1 km perimeter) is to avoid frequent handovers between close cell towers. However, on a large scale, such a precaution seems to have a negative impact. Furthermore, the extreme volatility of the performance of the time constraints criteria is remarkable. Clearly, this criterion is not able to cope with (changing) user behaviour during the summer months, resulting in the worst performance obtained.

## 5.3 The *35°-gap* in high-level validation

The most telling result of our analysis, is that all tested HDAs have CSM values which are still far off from the intended 0° as can be observed in Figure 2. Additionally, it is remarkable that CSM values for all HDAs occur in the same, rather small, range of CSM values (even though a small difference in CSM can induce a rather profound change in spatial patterns). The *35°-gap* observed is indicative for the (current) limits of single-step approaches based on simple decision rules, at least at cell tower level (as aggregation to higher levels might diminish the gap considerably).

The *35°-gap* also adds to the discussion of high-level validation. As mentioned before, the absence of individual-level validation is hindering a clear understanding of why the performance between algorithms may differ. Playing devil's advocate, one could care less about individual correctness as long as the statistical performance at nationwide level is sufficient. However, given the considerable differences between census and mobile phone home location data at cell tower level, as seen by both the *35°-gap,* together with the clear differences in spatial patterns as shown in Figure 3, it seems inevitable that investigations at individual or subset level will need to be undertaken to improve insights into the workings of HDA and, ultimately, the performance of home detection methods in general.

It is clear that the *35°-gap* requires further exploration so as to understand its constituent parts. We consider, at least, the following elements to be of importance:

- *Distorted local market shares*: Local market shares for individual mobile phone operators can be highly volatile and are often unknown. This causes a lot of uncertainty when it comes to high-level validation with census data, as the percentage of the population the operator actually captures in different regions is unclear. Unknown local market shares therefore impede both validation techniques that perform in pairs and/or absolute comparisons between population estimations and ground truth figures collected by nationwide censuses. They also most probably hinder validation techniques that are based on relative differences (like the CSM metrics).

- *Diversity of mobile phone use*: Differences in mobile phone use between users and/or regions can structurally influence the validation of single-step HDAs. When concentrated, differences in mobile phone use influence high-level validation in the same way as distorted market shares would. Additionally, it is clear that mobile phone usage changes rapidly over time. It can be argued that the use of phones for professional or private purposes was different in 2007 than it is today. Unfortunately, such usage contexts are not available in CDR data. Neither can they be easily derived since, in general, privacy regulations ban the linking of CDR data and customer databases that gather e.g. billing addresses or type of payment information. In other words, traces in CDR data will be of a different nature at different times because of differences in mobile phone usage. This implies that information on mobile phone usage is necessary to understand the effect on home detection performance.

- *Differing definitions of home*: Differences between the definition of home in census data and the definition of home by HDAs may cause structural discordance when validating the latter by the former. Even though official statistical practices have a tradition of distinguishing between different definitions of home, such as 'usual resident population' and 'second home population', it remains unclear to what degree mobile phone data is capable of capturing such concepts of home and to what degree different decision rules would favour the detection of different types of homes.

- *Technical aspects of the data collection and methodology*: Research has paid wide attention to the technical aspects of mobile phone data, especially when it comes to the estimation of cell tower areas and their translation into statistical areas (Ricciato et al. 2015). In our case, we avoided the translation problem by constructing a validation dataset at cell tower level, but for many cases this is not an option. Estimation of cell tower areas was done by Voronoi polygons, which introduces errors at a local scale but could also structurally influence high-level validation if, for example, areas of cell towers in high population density locations were consequently underestimated. Here too, the effects on high-level validation practices are currently unknown, but we expect them to be minor compared to previous points.

For all the points raised above, no quantification of their effect(s) has yet been explored. Additionally, it is worth nothing that some of these points may become more or less relevant over time due to, amongst others, any technical advancements or regulations. The EU General Data Protection Directive 2018, for example, will probably make it harder to work at the individual level (individual mobile phone use, different types of home definition). This will make high-level validation techniques more relevant, and thus increase the need to have a proper knowledge of local market shares. For this reason, it is difficult to assess the relevance of the findings given in this paper. However, we strongly suggest that all are the topics of future research and consideration.

# 6. Recommendations

Throughout this work, we have given suggestions concerning the use of HDAs for mobile phone data. In summary, we believe we can compile these suggestions into a set of three recommendations, which are relevant at different levels.

> **1. Individual level:** Currently, the biggest problem in ensuring the reliable use of HDAs for mobile phone data (and, in extension, other similar data sources like location based services or geo-tagged online social networks) is the absence of ground truth data at the individual level. We strongly recommend the collection of ground truth data linking mobile phone usage, the related CDR data, and movement patterns of individual users. Even if collected for only small samples of users, this step is essential to give proper estimations of error and performance of HDAs at individual level. It also would help in understanding the differences between decision rules, and plausibly allow for the development of non-generic HDAs that could switch between decision rules based on the characteristics of individual traces instead of being generically applied to all users. Additionally, the availability of individual level ground truth could shed light on the structural effects that currently obscure high-level validation practices, such as the changing usage of mobile phones and the differences between declared homes and lived-in homes as captured by census and mobile phone data respectively.
>
> **2. National level:** Apart from ensuring nationally representative sampling of individual level ground truth data, we believe it to be important either to understand local market shares of single operators, or to collect mobile phone data from all operators in the territory. Without this information, high-level validation of population estimations at nation-wide level will remain flawed, making it impossible to describe correctly the performance of HDAs at a larger geographical scale. In addition, resolving the local market share issue is a crucial step in the investigation of the (spatial) representativity of available mobile phone datasets, as unknown market shares at local level impede the analysis of subset populations in datasets.
>
> 3. **International level**: Finally, we believe that one of the key components to ensure a reliable use of mobile phone data in official statistics is the opportunity to test ideas and methodologies on different datasets, which contain differing populations and cover various time periods. This is not necessarily a matter of testing for uniformity. On the contrary, it is a matter of understanding the limits of current methodologies, assessing the true potential for applications and anticipating the wider challenges posed by fast-evolving technology usage and deployment. All of these factors are necessary to ensure the future applicability of mobile phone data sources in official statistics.

As we reflect on the direction further investigation should take, together with the feasibility of carrying out the recommendations proposed, we realize that this is a larger intervention than any single researcher, research group, national statistics office or even operator can be expected to take. Therefore, it is encouraging to see that collaborations are being formed to address different parts of the problem.

In France, for example, a collaboration between the operator Orange and the national statistics office INSEE is investigating different aspects of the high-level validation of home detection practices, such as translating Voronoi polygons into existing statistical grids (Sakarovitch et al. In Prep.).

On a European scale, the ESSnet Big Data project has been organising the exchange of best practices for the integration of mobile phone data (and multiple other big data sources) in official statistics. Its goal is directly in line with the recommendations previously described (especially recommendation 3: international level), facilitating the uniformity of quality and methodologies for the use of big data sources in European official statistics (ESSnet Big Data 2018).

As a last example, the Open Algorithm project (OPAL) is a collaboration between operators, academia, and institutional partners who are building a platform to allow the use of so-called *Open Algorithms* on mobile phone datasets from different operators (OPAL 2018). The idea is that users can launch a predefined set of algorithms (such as home detection algorithms), which are then run behind the firewalls of the operators before returning the aggregated results back to the user. Although the project is currently still in its test phase (with pilots in Senegal and Columbia), hopes are that it could facilitate cooperation between different operators in sharing basic statistical information from their datasets (as captured by the predefined set of algorithms). If all a country's mobile phone operators would engage in this form of cooperation, the problem of dealing with a distorted market share, for example, would be solved.

Hence, the bottom line is that although the home location problem is mainly a methodological one, the paths to address the problem are much more complex. They require the combination of collaborative, technical, methodological, institutional and strategic actions. Optimistically, we believe that official statistics offices are in a good position to (continue to) play a prominent role, because of their organisational structure, methodological knowledge and recognised institutional role within a country.

## 7. Conclusion

Big data sources in general, and mobile phone data in particular, create intriguing new opportunities and challenges for official statistics. Because of this, there has been a clear call for exploratory pilot projects to be carried out, as well as a trend towards critical investigation and transparency of methodologies to produce high-quality statistics. This paper adhered to both of these calls in its analysis of home detection practices for non-continuous location traces, focusing mainly on mobile phone data.

Based on a critical review of literature, we discussed how existing methods to identify home locations using non-continuous location traces mainly consist of single-step approaches that deploy simple decision rules and use high-level validation only.

We argued that, given the absence of ground truth data at individual level, i) it is unclear why one-step approaches are preferred over two-step approaches that are typically used for continuous location traces; ii) no consensus in literature exists on which criteria best to deploy when creating decision rules for home detection methods, nor has work been done to investigate the sensitivity of the results to these decision rules and criteria; and iii) the trustworthiness of high-level validation and its added value to the home detection practices are questionable at best.

By deploying five algorithms with simple decision rules to a large French CDR dataset, we demonstrated several of the problems. At individual level, we found home detection methods to be rather sensitive to criteria choice, with pair comparison of different home detection algorithms resulting in different identified homes for up to 40% of users. When looking at high-level validation, we found that five different home detection algorithms performed in a similar range (34°-38°) with a similar sensitivity to the time period and the duration for which the mobile phone data was collected. Even though we found that the sensitivity to time and the differences between different HDA algorithms does not seem large when expressed in CSM values, we showed how small changes to CSM values influence substantive and nationwide changes in the spatial patterns of population counts.

Our most noteworthy finding is the magnitude of the mismatch (the *35°-gap*) between population counts constructed from mobile phone-based data on home location and a validation dataset based on census data. This large mismatch is indicative of the severity of the home location problem and challenges the validity of single-step approaches in literature. In our discussion, we listed several elements that plausibly effect this mismatch but go unnoticed when only high-level validation is undertaken. We believe that these (structural) elements, such as unknown market shares and differences in mobile phone usage, need further investigation if ever home detection methodologies are to comply with official statistics' standards.

Finally, we compiled our findings, insights, and experiences into a set of specific recommendations, ranging from the collection of individual ground truth data to the testing of methods on multiple datasets. Given the nature of these recommendations and the tasks at hand, we think that it is unlikely that individual researchers, research groups, national statistics offices, or even mobile phone operators can, or will, invest in them. Therefore, we call on and support any ongoing, collaborative actions that tackle these problems, while recognising the prominent role official statistics can (continue to) play in this area.